\documentclass[pre,amsmath,superscriptaddress,preprint,nofootinbib]{revtex4}

\usepackage{graphicx}
\usepackage{bm}

\newcommand{\vecr}{\mathbf{r}}

\newcommand{\vecu}{\mathbf{u}}

\newcommand{\veca}{\mathbf{a}}

\newcommand{\vecF}{\mathbf{F}}

\newcommand{\vecf}{\mathbf{f}}

\newcommand{\lang}{\left\langle}

\newcommand{\rang}{\right\rangle}

\begin{document}

\title{Unbiased estimators for spatial distribution functions of classical fluids}

\author{Artur B. Adib}
\email{artur@brown.edu}
\affiliation{
Theoretical Division, T-13, MS B213, Los Alamos National Laboratory, Los Alamos, New Mexico 87545, USA
}
\affiliation{
Department of Physics, Box 1843, Brown University, Providence, Rhode Island 02912, USA
}

\author{Christopher Jarzynski}
\email{chrisj@lanl.gov}
\affiliation{
Theoretical Division, T-13, MS B213, Los Alamos National Laboratory, Los Alamos, New Mexico 87545, USA
}

\date{\today}

\begin{abstract}
We use a statistical-mechanical identity closely related to the familiar virial theorem,
to derive unbiased estimators for spatial distribution functions of classical fluids.
In particular, we obtain estimators for both the fluid density $\rho(r)$ in the vicinity of a fixed solute,
and for the pair correlation $g(r)$ of a homogeneous classical fluid.
We illustrate the utility of our estimators with numerical examples, which reveal advantages over
traditional histogram-based methods of computing such distributions.
\\ \\
LAUR-04-6270
\end{abstract}

\maketitle


\section{Introduction} \label{intro}

In molecular simulations of fluids, one is often interested in estimating a
{\em radial distribution function} $\rho(r)$, that is,
the average density of particles as a function of distance $r$ away from either a
tagged reference particle (the pair correlation function of uniform fluids\cite{frenkel02}),
or a fixed point in space (useful in the context of nonuniform fluids\cite{weeks02}).
Almost invariably, such estimates are constructed using histograms,
with $\rho(r)$ obtained by counting
particle visits to a spherical shell of radius $r$ and thickness 
$\Delta r$ centered about the point of interest \cite{frenkel02,chandler87}.
While simple to implement, this approach inevitably introduces systematic error,
or {\em bias},
as the intended estimator of $\rho(r)$ is in reality an estimator
of the particle density averaged over the finite thickness $\Delta r$ of the shell.
This bias can of course be made small by reducing $\Delta r$,
but only at the price of increased statistical error:
the smaller the bin size $\Delta r$, the fewer the particles found in a typical shell.

In this paper we show how to construct {\it unbiased} estimators of radial distribution
functions.
Upon ensemble averaging these estimators coincide with the desired observables,
thus eliminating the systematic error that arises in the histogram methods,
and altogether avoiding the problem of finite bin size.

Our estimators express a local, surface quantity in terms of a volume average;
for a given $r$, contributions to the estimate of $\rho(r)$ come from particles
throughout the bulk of the fluid,
and not only from those in a thin spherical shell at $r$.
This approach is similar to the widely used virial method for computing 
the pressure $p$ of a fluid\cite{becker67}:
while $p$ can be defined operationally as the force per unit area exerted by the fluid
on the walls of its container (a surface quantity),
the virial theorem re-expresses $p$
as a sum of the ideal gas pressure
and an exact correction given in terms of particle-particle forces throughout
the bulk of the fluid.

In what follows we first introduce the identity that underlies our approach
(Eq.\ref{uvirial}),
then we apply this identity to obtain estimators of commonly studied radial distribution
functions (Eqs.\ref{eq:rho_hat}, \ref{eq:rho_phi=U}, \ref{eq:rho_phi=hardSphere},
\ref{eq:gofr}, \ref{eq:rho_phi}).
We present numerical evidence illustrating the practical utility of our approach
in the context of a Lennard-Jones fluid,
and we conclude with a discussion of potential extensions applicable to more
elaborate situations.

\section{Theory}

Consider a fluid of $N$ identical, classical particles confined within a cubic
simulation box of dimensions $L\times L\times L$ and volume $V=L^3$,
with periodic boundary conditions.
\footnote{
As often the case when using periodic boundary conditions,
we assume that the interaction potential between pairs of particles has a finite range
no greater than $L/2$, so that each particle interacts with at most one image of
each other particle.}
Let $\Phi(\vecr_1,\ldots,\vecr_N)$ be the potential energy expressed in terms of the
locations of the particles.
This potential includes both particle-particle interactions and
nonuniform terms such as the interaction of each particle with a fixed solute.
The density $\rho(\vecr)$ in the canonical ensemble is given by:\cite{chandler87}
\begin{equation} 
\label{rho}
\rho(\vecr) = \frac{1}{Z_r} \sum_{i=1}^N \int_V d\vecr^{(-i)} 
\left. e^{-\beta \Phi(\vecr_1,\ldots,\vecr_N)} \right|_{\vecr_i = \vecr},
\end{equation}
where $\int_V d\vecr^{(-i)}$ denotes an integration over the system volume $V$ and 
all coordinates except $\vecr_i$,
and $Z_r \equiv \int d\vecr_1 \cdots d\vecr_N e^{-\beta \Phi}$ the configurational part 
of the partition function.
Let us now choose an arbitrary three-dimensional vector field $\vecu(\vecr)$, 
and apply the divergence (Gauss's) theorem to the product $\rho\vecu$:
\begin{equation} \label{start}
\oint_{\partial\Omega} \! d\veca \cdot \vecu \, \rho = 
\int_\Omega d\vecr \, \rho \nabla \cdot \vecu + \int_\Omega d\vecr\, \vecu \cdot \nabla \rho.
\end{equation}
The two terms on the right are integrals over an arbitrary volume $\Omega$
contained entirely within the simulation box;
the term on the left is an integral over the surface $\partial\Omega$ of this volume.
Eqs.\ref{rho} and \ref{start} combine to give
\begin{equation} 
\label{uvirial}
\oint_{\partial\Omega} \! d\veca \cdot \vecu \, \rho = 
\int_\Omega d\vecr \, \rho \nabla \cdot \vecu + 
\beta \lang \sum_{i \in \Omega} \vecu_i \cdot \vecF_i \rang,
\end{equation}
where the sum is over all particles inside the domain $\Omega$,
the notation $\vecu_i$ is shorthand for $\vecu(\vecr_i)$,
and $\vecF_i = - \nabla_i \Phi$ is the net instantaneous force on particle $i$. 
The angle brackets denote the canonical average,
$\lang \cdots \rang = Z_r^{-1} \int d\vecr_1 \cdots d\vecr_N e^{-\beta \Phi} \cdots \,$. 

In the examples discussed below,
the force $\vecF_i$ on a given fluid particle includes contributions from 
interactions with the other $(N-1)$ fluid particles, as well as a contribution
from a potential $U(\vecr)$, which represents a fixed
solute whose presence is felt by all $N$ fluid particles.
Since it is often convenient to distinguish between these two contributions,
we will use the notation $\vecf_i$ to denote the force on particle $i$
arising specifically from interactions with the other fluid particles,
and $-\nabla_i U(\vecr_i)$ to denote the force due to the solute;
thus, $\vecF_i = -\nabla_i U(\vecr_i) + \vecf_i$.

For the special choice $\vecu(\vecr)=\vecr$, Eq.\ref{uvirial} above reduces to
a result derived by Henderson (Eq.9 of Ref.\cite{henderson86}; see also Ref.\cite{powles85}),
which relates the density of particles on the surface $\partial\Omega$
to the density within the volume $\Omega$.
In that case the last term in Eq.\ref{uvirial} is very
similar to the familiar Clausius virial\cite{becker67}.
More generally, however, 
Eq.\ref{uvirial} encompasses a large family of virial-like 
theorems, each associated with a particular choice of $\vecu$. 
Such ``hypervirial'' relations have been studied by Hirschfelder \cite{hirschfelder60}, 
though they have been derived and expressed in a different way. 
We now show how, by choosing $\vecu(\vecr)$ to be discontinuous
over the surface of a sphere of radius $R$, 
we obtain useful identities for the fluid density averaged over this surface.

\subsection{Hard spherical solute}

To begin, let us consider a fluid in the presence of a fixed hard spherical solute.
This simple solvation model plays an important role in the study of 
nonuniform fluid theories
\cite{huang00,katsov01,weeks02}.
A quantity of particular interest is the {\em contact density}, $\rho_c$
(the particle density in the immediate vicinity of the hard sphere),
which is closely related to the free energetic cost of ``growing'' such a hard sphere
solute in the fluid\cite{weeks02}.

We will use a cartesian coordinate system whose origin $\vecr={\bf 0}$ is at the center of the box,
and we assume that this point also coincides with the center of the hard solute.
Let $R_{\rm sol}$ denote the radius of the solute,
$R_{\rm max}=L/2$ the radius of the largest sphere that fits inside the simulation box,
and $\Omega$ the thick spherical region between these two radii.
That is, $\Omega$ is defined by $R_{\rm sol} < r < R_{\rm max}$,
and the volume of this region is
$V_\Omega = (4\pi/3)(R_{\rm max}^3-R_{\rm sol}^3)$.
With the aim of determining the particle density  at a distance $R$ from 
the center of the solute (where $R_{\rm sol} < R < R_{\rm max}$), let us construct
the vector field
\begin{equation} 
\label{u-density}
  \vecu(\vecr; R) = 
      \begin{cases}
         \dfrac{r^3 - R_{\rm sol}^3}{3 r^2} \, \hat\vecr, & {(R_{\rm sol} \leq r \leq R)} \\ \\
         \dfrac{r^3 - R_{\rm max}^3}{3 r^2} \, \hat\vecr, & {(R < r \leq R_{\rm max})}
      \end{cases}
\end{equation}
where the notation $\vecu(\vecr;R)$ indicates the parametric dependence of $\vecu$ 
on the radius of interest $R$.
This field vanishes at the boundaries of $\Omega$ (i.e.\ at $R_{\rm sol}$ 
and $R_{\rm max}$), therefore so does the term on the left side of Eq.\ref{uvirial}.
The divergence of $\vecu$ inside the region $\Omega$ is easily calculated to be
\begin{equation} \label{divu}
  \nabla \cdot \vecu =  1 - \dfrac{V_\Omega}{4\pi R^2} \,\delta(r-R),
\end{equation}
where the delta-function arises from the discontinuity of ${\vecu}$
at the surface $\vert{\vecr}\vert=R$.
From this we get
\begin{equation}
\int_\Omega d\vecr \, \rho \nabla \cdot \vecu   =
\lang N_\Omega \rang - V_\Omega \, \rho(R),
\end{equation}
where $\lang N_\Omega \rang = \int_\Omega d\vecr \, \rho(\vecr)$ is the average
number of particles found in the region $\Omega$,
and 
\begin{equation}
\rho(R) = \frac{1}{4\pi R^2}\int d\vecr \,\rho(\vecr)\,\delta(r-R)
\end{equation}
is the particle density, averaged over the spherical surface $r=R$.
Combining this result with Eq.\ref{uvirial} and rearranging terms finally gives us:
\begin{equation} \label{rho_final}
  \rho(R) = \frac{1}{V_\Omega}
                     \Biggl[
                     \lang N_\Omega \rang + 
                     \beta \lang \sum_{i\in \Omega} \vecu_i
                     \cdot \vecf_i \rang
                     \Biggr],
\end{equation}
where $\vecu_i = \vecu(\vecr_i;R)$.
Note that we have replaced $\vecF_i$ in Eq.\ref{uvirial} by $\vecf_i$ above;
because the region $\Omega$ ($R_{\rm sol}<r<R_{\rm max}$) was defined to be entirely 
outside the region of space occupied by the solute ($r\le R_{\rm sol}$),
the net force felt by any particle
that is inside $\Omega$ includes only contributions from the other fluid
particles.
Eq.\ref{rho_final} is an {\it exact} result that expresses the surface-averaged density $\rho(R)$ 
in terms of a bulk average: all the particles inside $\Omega$ contribute to the right side
of Eq.\ref{rho_final}.

As a consistency check, we note that
if our fluid is an ideal gas, then $\vecf_i$ is identically zero, and Eq.\ref{rho_final}
becomes $\rho(R) = \lang N_\Omega\rang/V_\Omega$.
This is as expected:
in the ideal gas limit the fluid density is independent of the distance
away from the solute, and is therefore equal to the average number of particles
in any given region of space, divided by the volume of that region.
Thus for the more general case of a non-ideal fluid,
the two terms inside the square brackets in Eq.\ref{rho_final} can be interpreted as,
respectively,
an ideal gas contribution to the radial distribution function,
and a virial-like correction due to interparticle interactions.

Eq.\ref{rho_final} is equivalent to the statement that the quantity
\begin{equation}
\label{eq:rho_hat}
\hat\rho(\Gamma;R) = \frac{1}{V_\Omega}
                     \Biggl(
                     N_\Omega + 
                     \beta \sum_{i\in \Omega} \vecu_i
                     \cdot \vecf_i 
                     \Biggr)
\end{equation}
is an {\em unbiased estimator} of the radial distribution function $\rho(R)$.
Here, $\Gamma=(\vecr_1,\vecr_2,\cdots,\vecr_N)$ is a single N-particle configuration
of our system,
and the quantity on the right side above is well defined for any such configuration.
Thus, $N_\Omega = N_\Omega(\Gamma)$ is the number of fluid particles inside $\Omega$,
for the configuration $\Gamma$.
By the term ``unbiased estimator'', we mean simply that the average of
$\hat\rho(\Gamma;R)$, over configurations $\Gamma$ sampled from an equilibrium (canonical)
ensemble, is identically equal to the desired density $\rho(R)$;
that is, $\rho(R) = \lang \hat\rho(\Gamma;R) \rang$.
This is the content of Eq.\ref{rho_final}.

In the example that we have just presented,
${\vecu}({\vecr};R)$ was chosen so as to 
(a) cause the surface term in Eq.~(\ref{uvirial}) to vanish, and 
(b) ``pick out'' the density at a particular distance $R$ from the origin via a discontinuity at $R$. 
We now show how the same strategy can be applied when the solute is no longer perfectly hard. 
This will pave the way to the consideration of particle-particle distribution functions in 
homogeneous fluids.

\subsection{General spherical solute}

Consider a solute described by a fixed spherically symmetric potential $U(r)$.
Recalling that $\vecF_i = -\nabla_i U(r_i) + \vecf_i$,
Eq.~(\ref{uvirial}) can be rewritten as
\begin{equation}
\label{uvirial-extern}
  \oint_{\partial\Omega} \! d\veca \cdot \vecu \, \rho = 
  \int_\Omega d\vecr \, \rho \,
  \hat{\cal L}\cdot\vecu
        + \beta \lang \sum_{i \in \Omega} \vecu_i \cdot \vecf_i \rang,
\end{equation}
where
\begin{equation}
\label{eq:defL}
\hat{\cal L}\cdot\vecu \equiv
e^{\beta U} \nabla \cdot ( e^{-\beta U} \vecu ).
\end{equation}
Essentially, the solute contribution to $\vecF_i$ in the second term on the right side
of Eq.\ref{uvirial}, has been transferred to the first term on the right.

Let us now take $\Omega$ to be the entire region of space enclosed by a sphere 
of radius $R_{\rm max} = L/2$.
By analogy with the hard sphere example, let us search for a vector
field $\vecu(\vecr;R)$ that:
(a) vanishes on the surface of $\Omega$,
(b) is discontinuous at $r=R$, and
(c) is spherically symmetric and oriented along the radial direction,
i.e.\ $\vecu(\vecr;R) = u(r;R)\,\hat\vecr$.
We also want $u(0;R)=0$, to avoid a singularity at the origin.
Note that these properties do not {\it uniquely} specify $\vecu(\vecr;R)$. 
In the following two paragraphs we will construct two different fields satisfying 
these conditions (Eqs.\ref{eq:u_phi=U} and \ref{eq:u_phi=hardSphere})
and we will present the unbiased estimators associated with these fields
(Eqs.\ref{eq:rho_phi=U} and \ref{eq:rho_phi=hardSphere}).
In the Appendix, we show how this approach can be generalized.

With Eq.\ref{divu} in mind, let us first search for a field $\vecu(\vecr;R)$
that satisfies
\begin{equation}
\label{eq:Ldotu}
\hat{\cal L} \cdot\vecu = 1 - \alpha \delta (r-R),
\end{equation}
where $\alpha$ is a constant whose value is determined by demanding that 
$\vecu$ vanish at the origin and at the boundary of $\Omega$ ($r=R_{\rm max}$).
The solution of Eq.\ref{eq:Ldotu} is obtained by straightforward integration:
\begin{equation} \label{eq:u_phi=U}
  \vecu = \hat\vecr \frac{e^{\beta U(r)}}{r^2} \times
      \begin{cases}
         \int_0^r ds\, s^2 e^{-\beta U(s)}, & 0\leq r \leq R \\ \\
         \int_{R_{\rm max}}^r ds\, s^2 e^{-\beta U(s)}. & R < r \leq R_{\rm max}
      \end{cases}.
\end{equation}
This field clearly satisfies the desired boundary conditions, and by explicit
evaluation we can confirm that
\begin{equation}
\label{L1}
\hat{\cal L}\cdot\vecu =
1 - 
\delta(r-R)
\frac{e^{\beta U(R)}}{4\pi R^2}
Q_U,
\end{equation}
where
\begin{equation}
Q_U = 4\pi \int_0^{R_{\rm max}} ds\,s^2 e^{-\beta U(s)}.
\end{equation}
Combining Eqs.\ref{uvirial-extern} and \ref{L1} and carrying out the sort of
rearrangement of terms that led to Eqs.\ref{rho_final} and \ref{eq:rho_hat},
we arrive at the following unbiased estimator for $\rho(R)$:
\begin{equation}
  \label{eq:rho_phi=U}
  \hat \rho(\Gamma;R) = \frac{e^{-\beta U(R)}}{Q_U}\left(
     N_\Omega + 
     \beta \sum_{i \in \Omega} \vecu_i \cdot \vecf_i \right).
\end{equation}
This is very similar to Eq.\ref{eq:rho_hat}.
In particular, the two terms inside the square brackets represent an ideal
gas contribution to $\rho(R)$, and a virial-like correction.
As in the example of the hard-sphere solute, $\hat\rho(\Gamma;R)$ takes on different
values for different configurations $\Gamma$, but its average over an equilibrium
ensemble of configurations is equal to $\rho(R)$.

Although Eq.\ref{eq:rho_phi=U} is correct, the evaluation of the function 
$\vecu(\vecr;R)$ involves an integral of the form
$\int^r ds\,s^2 e^{-\beta U(s)}$ (see Eq.\ref{eq:u_phi=U}).
In an actual application of this method this integral must generally
be carried out numerically.
We can avoid this inconvenience by replacing the integral in Eq.\ref{eq:u_phi=U}
by an analytically tractable expression.
For instance, if the solute is strongly repulsive within a region $r<R_1$,
then we can replace the factor $e^{-\beta U(s)}$ appearing in the integrand,
by the unit step function $\theta(s-R_1)$.
We then get:
\begin{eqnarray}
\label{eq:u_phi=hardSphere}
  \vecu(\vecr; R) &=&  \hat\vecr \dfrac{e^{\beta U(r)}}{3r^2} 
  \Bigl[
  (r^3-R_1^3)\theta(r-R_1) + (R_1^3-R_{\rm max}^3)\theta(r-R)
  \Bigr]
\\
\hat{\cal L}\cdot\vecu &=& e^{\beta U(r)}
\left[
\theta(r-R_1) - \frac{\delta(r-R)}{4\pi R^2} V_{\Omega^*}
\right],
\\
\label{eq:rho_phi=hardSphere}
  \hat\rho(\Gamma;R) &=& \frac{e^{-\beta U(R)}}{V_{\Omega^*}}
  \left[
  \sum_{i\in \Omega^*} e^{\beta U(r_i)} 
  + \beta \sum_{ i \in \Omega} \vecu_i \cdot \vecf_i
  \right],
\end{eqnarray}
where 
$\Omega^*$ is the spherical shell defined by $R_1<r<R_{\rm max}$,
whose volume is 
$V_{\Omega^*} = (4\pi/3)(R_{\rm max}^3-R_1^3)$.
This field satisfies the desired conditions listed two paragraphs ago, 
but the evaluation of $\vecu(\vecr;R)$ in this case
no longer requires numerical integration.

Note that the replacement of $e^{-\beta U(s)}$ (in Eq.\ref{eq:u_phi=U})
by $\theta(s-R)$ (to get Eq.\ref{eq:u_phi=hardSphere})
simply amounts to a different choice of $\vecu(\vecr;R)$.
The resulting estimator, Eq.\ref{eq:rho_phi=hardSphere}, is as exactly
unbiased as the one given by Eq.\ref{eq:rho_phi=U}.
In the Appendix we show that, in fact, both these estimators are special
cases of a more general expression for $\hat\rho(\Gamma;R)$ (Eq.\ref{eq:rho_phi}).

\subsection{Pair correlation function}

We now consider how our method can be applied to the estimation of the pair
correlation function $g(r)$ of a uniform fluid.
As above, we imagine $N$ identical, classical particles inside a cubic simulation
box of volume $V=L^3$ and periodic boundary conditions, only now we assume that 
there are no solutes present.
The total potential energy of a given configuration of paricles is given by
$\sum_{i<j} v(r_{ij})$,
where $v(r)$ is the pairwise interaction potential (e.g.\ Lennard-Jones),
$r_{ij}$ is the distance between particles $i$ and $j$, and
the sum is over all pairs of particles.

For purpose of presentation, let us choose one of our $N$
particles to act as a reference particle, and imagine for the moment
that we hold 
the position of that particle fixed at the center of the simulation box.
The pair correlation function is given by $g(r) = \rho(r)/\rho_b$,
where $\rho(r)$ is the density of particles at a
distance $r$ from the reference particle,
and $\rho_b=(N-1)/V$ is the bulk density of the unconstrained particles
(see e.g.\  Ref.~\cite{chandler87}, pp. 196-197.)
We can think of the reference particle as a fixed, spherically symmetric solute,
which happens to be physically identical to the remaining fluid particles.
Since Eqs.\ref{eq:rho_phi=U} and \ref{eq:rho_phi=hardSphere} both provide
unbiased estimators for the density profile around such a solute, we can apply
either of those results to the present case, replacing $U(r)$ by $v(r)$ to get
an unbiased estimator for $g(R)$.
For instance, from Eq.\ref{eq:rho_phi=hardSphere} we get
\begin{equation}
\label{eq:gofr}
  \hat g(\Gamma;R) = 
  \frac{e^{-\beta v(R)}}{\rho_b V_{\Omega^*}}
  \left[
  {\sum_{i\in \Omega^*}}^\prime e^{\beta v(r_i)} 
  + \beta {\sum_{ i \in \Omega}}^\prime \vecu_i \cdot \vecf_i^\prime
  \right].
\end{equation}
Here the notation
$\sum^\prime$ indicates that the reference particle is not included in the sum
(since that particle is being treated as a fixed solute),
and similarly $\vecf_i^\prime$ is the net force acting on particle $i$ from all other particles
except the reference particle.
The $\vecu$-field that enters Eq.\ref{eq:gofr} is given by Eq.\ref{eq:u_phi=hardSphere},
with $U(r)$ replaced by $v(r)$.

The discussion of the previous paragraph was framed around the idea that a single
reference particle is held fixed, and $g(R)$ is computed with respect to that particle
alone.
However, in an actual implementation of the method, all $N$ particles would be
treated equally during the simulation.
For every configuration $\Gamma=(\vecr_1,\cdots,\vecr_N)$ generated during such a simulation,
the right side of Eq.\ref{eq:gofr} would be evaluated $N$ times, with each particle
taking its turn as the reference particle.
\footnote{The vector $\vecr_i$ then refers to the displacement of the $i$'th particle relative
to the current reference particle,
i.e.\ we temporarily treat the location of the reference particle as the origin
of the coordinate system.}
Each of these $N$ evaluations represents a single unbiased estimate of $g(R)$;
by averaging over all $N$ of them, we
obtain better statistics than by the use of only a single reference particle.
Effectively, the same procedure is used when applying the usual histogram method:
instead of specifying one reference particle and compiling
statistics on the relative locations of the other $N-1$ particles,
the pair correlation function $g$ is estimated by keeping track of all
$N(N-1)/2$ pairwise distances in the fluid.

\subsection{Generalization to $NPT$ ensemble}

So far in our presentation, we have taken the volume $V$ of the simulation box to be
constant, as appropriate for quantities computed in the $NVT$ equilibrium ensemble.
However, our method easily extends to the $NPT$ ensemble, which can be viewed as
a superposition of $NVT$ ensembles.
Explicitly, in the $NPT$ case we have
\begin{equation}
\rho(R) = \lang\lang \hat\rho(\Gamma;R,V) \rang\rang =
\lim_{M\rightarrow\infty} \frac{1}{M}
\sum_{m=1}^M \hat\rho(\Gamma_m;R,V_m).
\end{equation}
The double angle brackets signify an average over both a distribution of system volumes $V$,
and for each volume a canonical average over microstates $\Gamma$ sampled at that fixed volume.
The quantity $\hat\rho(\Gamma;R,V)$ is any one of the unbiased estimators derived above,
with the dependence on $V$ (or equivalently, $L$) made notationally explicit.
The final term in the equation above refers to an average over $M$ configurations
generated during an $NPT$ simulation, in which both the microstate $\Gamma$ and the volume $V$
fluctuate with time.

\section{Numerical results}
\label{sec:numres}

\begin{figure*}
\begin{center}

\includegraphics[width=3.5in,angle=270]{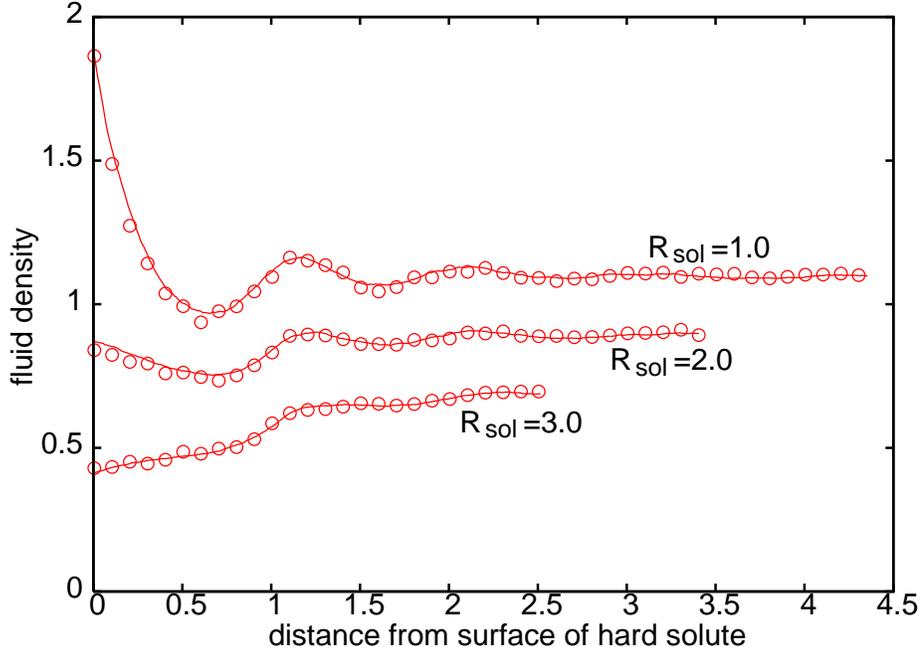}
\caption{\label{proof_of_principle}
Estimates of the fluid density obtained from the histogram method (circles)
and with the unbiased estimator of Eq.\ref{eq:rho_hat} (lines).
For the histogram method, we used a bin size $\Delta r=0.01$, although only every
tenth bin is shown in the figure, to avoid cluttering.
When using our unbiased estimator, $\rho(r)$ was evaluated at increments of
$0.01$, resulting in the smooth curves shown above.
Finally, the curves corresponding to $R_{\rm sol}=2.0$ and $R_{\rm sol}=1.0$ have
been shifted upward by $0.2$ and $0.4$ units, respectively, again to avoid
cluttering.
}
\end{center}
\end{figure*}

To test our method, we have performed Monte Carlo simulations of a truncated and
shifted Lennard-Jones (LJ) fluid both in the $NVT$ and $NPT$ ensembles \cite{frenkel02}. 
We have checked the reliability of our simulations with the available literature data on 
the LJ equation of state \cite{johnson93}, always finding good agreement within
statistical error bars estimated through the block average method 
of Flyvjerg and Petersen \cite{flyvbjerg89,frenkel02}.

As a proof-of-principle test case, we first ran three simulations of 
a Lennard-Jones fluid surrounding a fixed hard spherical solute, in the $NPT$ ensemble,
with number of particles $N=864$, temperature $T=0.85$, and pressure $p=0.022$.
We used a Lennard-Jones cut-off radius $r_c=2.5$ for the fluid particles, and
the solute radius was taken as $R_{\rm sol}=1.0$, $2.0$, and $3.0$ for the three simulations.
(Here and below we use the reduced LJ units \cite{frenkel02} when specifying these
parameters.)
These values are identical to those used to produce Fig.3 of Huang and Chandler\cite{huang00}.
From the data generated during each simulation ($2^{13}$ statistically independent configurations),
we estimated the radial distribution
function $\rho(r)$ using the histogram method, as well as with the unbiased estimator
given by Eq.\ref{eq:rho_hat} above (with $\vecu$ given by Eq.\ref{u-density}).
Fig.\ref{proof_of_principle} shows good agreement between the two.

\begin{figure}
\begin{center}

\includegraphics[width=3.5in,angle=270]{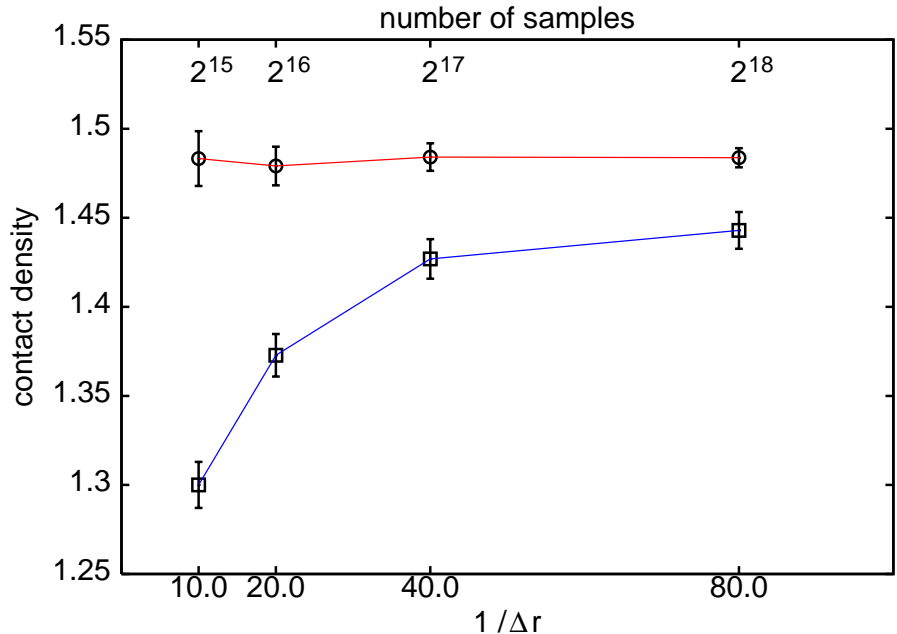} 
\caption{\label{contact_density} 
Estimates of contact density at the surface of a hard cavity of radius $R_{\rm sol}=1.0$,
obtained from four separate simulations.
The lower points (squares) were obtained with the histogram method.
The upper points (circles) were obtained using Eq.\ref{eq:rho_hat}.
(The connecting lines are guides to the eyes.)
Each simulation is identified by the number of samples generated,
and by the inverse bin size used in the estimates based on the histogram method.
}
\end{center}
\end{figure}

\begin{figure}
\begin{center}

\includegraphics[width=3.5in,angle=270]{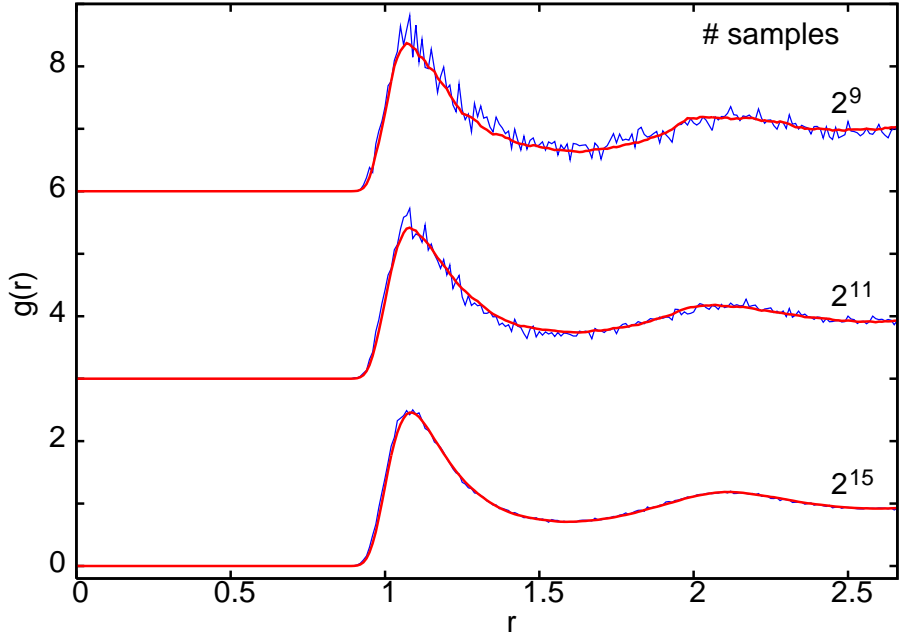}
\caption{\label{gofr} The particle-particle distribution function $g(r)$ for a 
uniform Lennard-Jones fluid, comparing the present method (thick red lines) 
with the usual histogram (thin, noisy blue lines) for different numbers of samples.
Each curve was evaluated at points spaced by $\Delta r=0.01$,
corresponding to the bin size used in the histogram method.
The curves have been shifted vertically in order to facilitate visualization.
The error bars of the present method (not shown) are always smaller than 
the histogram ones.}
\end{center}
\end{figure}

Next, to highlight the bias due to finite bin size when using the histogram method
-- and to illustrate that our method does not suffer from this bias --
we ran four new simulations of a solvated sphere of radius $R_{\rm sol}=1.0$,
with the $NPT$ parameters described in the previous paragraph.
The four simulations differed in duration, generating
$n_s=2^{15}$, $2^{16}$, $2^{17}$, and $2^{18}$ samples, respectively.
From each simulation we estimated the contact density $\rho_c = \rho(R_{\rm sol}^+)$
using both the histogram method 
(i.e.\ by counting the number of particles
within a distance $\Delta r$ from the surface of the sphere)
and the unbiased estimator of Eq.\ref{eq:rho_hat}.
With the histogram method, we varied the bin size $\Delta r$ from one simulation to the next,
keeping the product $n_s\,\Delta r$ constant
so that the number of counts in the bin was roughly the same for each simulation.
Because $\rho(r)$ increases sharply as one approaches the solute surface
(see the upper curve in Fig.\ref{proof_of_principle}), we expect the histogram method
to underestimate $\rho_c$ for any finite bin size,
but this systematic error should diminish as we decrease the size of the bin.
Fig.\ref{contact_density} shows the estimates of $\rho_c$ that we obtained from
our simulations.
The results obtained with the histogram method (squares) are consistent with our
expectations:
with decreasing bin size, the estimate of $\rho_c$ increases, suggesting an asymptotic
approach to the correct value as $\Delta r\rightarrow 0$.
However, this comes at a price:
more samples are needed to maintain the same level of statistical error.
When using Eq.\ref{eq:rho_hat} (circles), there is no discernable bias;
the statistical error bars simply become smaller as the number of samples increases.

Finally, Fig.\ref{gofr} shows estimates of the pair correlation function $g(r)$
obtained from the simulation of a uniform
Lennard-Jones fluid (no solute) in the $NVT$ ensemble, with
$N=108$, $V=N/\rho_b$ (where the bulk density was set at $\rho_b=0.70$), and $T=0.85$.
Estimates of $g(r)$ were obtained after the generation of
$2^9$, $2^{11}$, and finally $2^{15}$ samples,
using both Eq.\ref{eq:gofr} and the histogram method.
Although exactly the same data were used for each overlying pair of curves,
our unbiased estimator furnishes a smoother estimate of $g(r)$,
with the difference between the two methods diminishing with increasing
number of samples.

\section{Conclusions}

We have introduced and illustrated a general method for the numerical estimation
of spatial distribution functions in classical fluids. 
Our method relies on the construction of unbiased estimators
(see Eqs.\ref{eq:rho_hat}, \ref{eq:rho_phi=U}, \ref{eq:rho_phi=hardSphere},
\ref{eq:gofr}, \ref{eq:rho_phi}),
derived from a virial-like identity (Eq.\ref{uvirial}).
We have focused in this paper on cases for which the spatial distribution function
of interest has spherical symmetry, reflecting e.g.\ the symmetry of the
fixed solute,
but the extension of our method to cases of oddly shaped solutes,
or nonuniform averages over possibly non-spherical surfaces $\partial\Omega$,
is not as difficult as might first appear to be the case\footnote{
We would like to thank Hank Ashbaugh for valuable suggestions with regard
to this point.},
and we hope to study this issue in greater detail in future work.
Another natural extension would be in the direction of complex fluids.

While the use of simple histograms does not normally
pose a serious computational challenge to the estimation of spatial distribution
functions (at least for simple fluids),
the numerical results of Section \ref{sec:numres} suggest that our method
has certain advantages,
such as the elimination of systematic bias and the generation of smooth results.

\acknowledgments

The authors would like to thank Hank Ashbaugh, Gerhard Hummer, Lawrence Pratt, 
John Weeks, and Tom Woolf for fruitful discussions and suggestions.
This research was
supported by  the Department of Energy,  under contract W-7405-ENG-36.

\section*{Appendix}

Here we briefly outline a generalization of the approach that led to
Eqs.\ref{eq:rho_phi=U} and \ref{eq:rho_phi=hardSphere}.
The physical context is that of a fluid surrounding a spherically symmetric
solute, but as illustrated in the main text any results obtained in this
case can just as well be applied to the estimation of the pair correlation
function $g(r)$.

Recall that we want to find a $\vecu$ field that:
vanishes at $r=R_{\rm max}$;
is discontinuous at $r=R$;
and has the form $\vecu(\vecr;R)=u(r;R)\hat\vecr$, with $u(0;R)=0$.
These conditions are satisfied quite generally by the following field:
\begin{equation}
\label{eq:u-phi-general}
\vecu(\vecr;R) = \hat\vecr \frac{e^{\beta U(r)}}{r^2}
\int_{a(r)}^r ds\,s^2\,e^{-\beta\phi(s)}
\qquad,\qquad
a(r) = R_{\rm max} \cdot \theta(r-R),
\end{equation}
where $\phi(r)$ is an arbitrary function,
and $\theta(\cdot)$ is the unit step function.
Because of the way that $a(r)$ is defined,
$\vecu(\vecr;R)$ points {\em outward} (i.e.\ away from the
origin) for $r<R$, and {\em inward} for $r>R$,
as was the case with the fields defined by Eqs.\ref{u-density}, \ref{eq:u_phi=U}
and \ref{eq:u_phi=hardSphere}.
From Eq.\ref{eq:u-phi-general} we get
\begin{equation}
\label{Lu}
\hat{\cal L}\cdot\vecu = e^{\beta U(r)}
\left[
e^{-\beta\phi(r)} - \frac{\delta(r-R)}{4\pi R^2} Q_\phi
\right],
\end{equation}
where $Q_\phi = 4\pi\int_0^{R_{\rm max}} ds\,s^2e^{-\beta\phi(s)}$,
and $\hat{\cal L}$ is defined by Eq.\ref{eq:defL}.
Combining Eqs.\ref{uvirial-extern} and \ref{Lu} and once again rearranging
terms in the manner that led to Eqs.\ref{rho_final} and \ref{eq:rho_hat}, 
we get the following unbiased estimator for $\rho(R)$:
\begin{equation} 
  \label{eq:rho_phi}
  \hat\rho(\Gamma;R) = \frac{e^{-\beta U(R)}}{Q_\phi}\left[ 
     \sum_{i\in\Omega} e^{\beta(U-\phi)_i} + 
     \beta \sum_{i \in \Omega} \vecu_i \cdot \vecf_i \right],   
\end{equation}
where $(U-\phi)_i\equiv U(r_i)-\phi(r_i)$.

The function $\phi(r)$ plays the role of a free
parameter, to be chosen so as to maximize computational convenience.
If we choose $\phi(r)=U(r)$, then we arrive at the estimator
defined by Eqs.\ref{eq:u_phi=U} and \ref{eq:rho_phi=U}.
Alternatively, we can take $\phi(r)$ to be a hard spherical potential 
($\phi=+\infty$ for $r\le R_1$ and $\phi=0$ for $r>R_1$)
in which case we are led to the results given by
Eqs.\ref{eq:u_phi=hardSphere} and \ref{eq:rho_phi=hardSphere}.

\bibliography{mybib}

\end{document}